\documentclass[a4paper,12pt,twoside]{article} 

 \usepackage{ragged2e}
\usepackage{hhline}
 \usepackage{graphicx}
\usepackage{adjustbox}
\usepackage[numbers, square, comma, sort&compress]{natbib}
\usepackage{notoccite}
 \usepackage{float}
 \usepackage{amsmath}
\usepackage{rotating}
 \usepackage{mathtools}
\usepackage{hyperref} 
\usepackage{authblk}
\usepackage{footmisc}
\usepackage{resizegather}    
\usepackage{latexsym}
\usepackage{amssymb,epsf}
\usepackage{titling}
\usepackage{geometry}
\usepackage{float}

 \setcitestyle{aysep={}}  

\begin{document}

  \begin{center}

	{\Large {\bf On An Interacting New Holographic Dark Energy Model: observational constraints} }
	
	\vspace*{12mm}
	{\large  Ehsan Sadri$^{a,}$\footnote{ehsan@sadri.id.ir}, Martiros Khurshudyan$^{b,c,d,e,f,}$\footnote{khurshudyan@yandex.com, khurshudyan@ustc.edu.cn, khurshudyan@tusur.ru} }
	\vspace*{0.5cm}
	
	{ {\it $^a$Department of Physics, Central Tehran Branch, Islamic Azad University, Tehran, Iran}} \\
	{ {\it $^b$CAS Key Laboratory for Research in Galaxies and Cosmology, Department of Astronomy, University of Science and Technology of China, Hefei 230026, P. R. China }} \\
	{ {\it $^c$School of Astronomy and Space Science University of Science and Technology of China, Hefei 230026, P. R. China}} \\
	{ {\it $^d$Institut de Ciencies de lEspai (CSIC), Campus UAB, Carrer de Can Magrans, s/n 08193 Cerdanyola del Valles, Barcelona, Spain}} \\
	{ {\it $^e$International Laboratory for Theoretical Cosmology, Tomsk State University of Control Systems and Radioelectronics, 634050 Tomsk, Russia}} \\
	{ {\it $^f$Research Division, Tomsk State Pedagogical University, 634061 Tomsk, Russia}}
	
\vspace{15mm}

\end{center}

  \begin{abstract}
 In this paper, we study the interacting DGP braneworld Holographic Dark Energy model in a spatially flat FRW universe. Mainly, in this study we concentrate our attention on both interacting and non-interacting form of the model. The study shows that the equation of state and the deceleration parameter depict an accelerated universe for all variety of interactions. On the other hand, the StateFinder analysis shows that of the interacting and non-interacting behave similar to both quintessence and phantom dark energy and for the present value obey the behavior of quintessence. Moreover, the result of $Om$-diagnostic is an emphasis on the result of the equation of state showing that the current model is in the quintessence are with Phantom-like behavior in the late time. By the use of the squared sound speed $v^2_s$ we find that the present mode has a good stability. In order to obtain the best fit values of the parameters in this work we used the latest observational data (Pantheon, Boss DR12 and Planck 2015) implementing MCMC method by the use of EMCEE python package. We also employ Akaike Information Criterion (AIC) and Bayesian Information Criterion (BIC) model selection tools and compare the model with $\Lambda$CDM as the reference model.
  \end{abstract}
   \section{ INTRODUCTION}
  \justify
Dark energy - raised in 1998 \cite{101}- has become one of the main issues in modern cosmology
and many models have been proposed to investigate this new concept of cosmology. In spite of many proposed models, dark energy still remains one of the open issues in cosmology\cite{{14},{15},{16},{56},{64},{65},{66},{67},{69}} (to mention a few). 
 The cosmological constant $\Lambda$ due to its convenient interpretation of the universe's expansion can be considered as the good case for study of the dark energy\citep{{56},{1.2},{1.3},{101}}. Despite this appropriateness, the cosmological constant contains some drawbacks such inability to explain, why the densities of dark sectors (dark energy and dark matter) are of the same order since they evolve in distinct way is of these drawbacks\citep{{1.5},{1.6},{1.7},{1.8},{67}}. Thus, for alleviating these problems the holographic dark energy (HDE) as an alternative has been proposed and drawn many attentions in recent two decades \cite{{8},{9},{10},{11}}. This model is stemmed from the holographic principle related to which all of the information in a specific area of space can be drawn out from its boundary region and are constrained by an IR cutoff\cite{{12},{13}}. The energy density of HDE can be shown as $\rho_D=3c^2M_P^2/L^2$ \cite{{17},{18},{19}}. Using HDE compared to other models is more appropriate to investigate the problems of dark energy\cite{{13},{71},{72},{17},{74},{75}}. Studying of HDE, also help to avoid the formation of Black Holes which can be investigated by paying attention to values of the equation of state (less than -1) \cite{{52},{53},{54},{55},{87},{77},{24},{78}}. This issue was discussed also by Nojiri and Odintsov(and their collaborators) who checked the possibility and probability of the universe having an equation of state with phantom behavior\cite{{87}, {78}, {89}, {102},{103}}. On the other hand, the profound relation between the gravitational terms which describe in the bulk and the first law of thermodynamic can lead to different ideas of holography\cite{29}. In
recent years, brane theories embedded in a higher dimensional space-time has attracted more attention \cite{{79},{91},{92},{93}}. In these theories the cosmic evolution is explained by a Friedmann equation interacting with the bulk's effects onto the brane. The most popular model in the framework of braneworld has proposed as DGP which stands for DvaliGabadadze-Porrati \cite{85}. In DGP model the four dimensional universe is turned to five dimensional Minkowskian bulk. The self-accelerating characteristic of DGP model is able to convey the late time cosmic speed up without relation to
dark energy \cite{62},\cite{86}. This characteristic of DGP model also cannot satisfy the phantom line crossing and for this issue adding an energy feature on the brane is required \cite{29}. Regarding
this, an added dark energy component to the brane models lead to emergence of a novel way of explanation for late time acceleration and also better compatibility with observational points
\cite{29}. To check the usability of various models in the context of different cosmological frameworks, one can study the types of evolution and also behavior of the models under the accurate conditions. Despite that the evolution of cosmic expansion defined by Hubble parameter ($H$) and the rate of acceleration and deceleration of this expansion are defined by $q$ and $\omega_D$, we are not able intelligibly to identify variety of dark energy models by the use of these two parameters since for all cases $H > 0$ or $q < 0$. Hence, in order to have accurate calculations about this issue and due to the development in observational data during the recent two decades a new geometrical diagnostic pair-known as the StateFinder pair- for tracking the dark energy models has been proposed \cite{{31},{32}}.
\begin{equation}\label{r-s}
r=\frac{\dddot{a}}{aH^3}=1+\frac{\ddot{H}}{H^3}+3\frac{\dot{H}}{H^2}~~~~~~~~and~~~~~~~~s=\frac{r-1}{3\left(q-\frac{1}{2}\right)}.
   \end{equation}
This tool opens a new way to specify the features of dark energy and check the distance from the main HDE models. By the use of this advantageous tool, cosmologists trace the path of current models.\\
To find out the behavior of the dark energy models, also one can use $Om$-diagnostic tool. The $Om$-diagnostic tool due to its dependency on expansion rate specifies more easier from observations than StateFinder pair \cite{42}. The plot of this tool has two parts: Phantom-like part for positive trajectories and quintessence for negative trajectories. The $Om$-diagnostic term can be written as
 \begin{equation}\label{om}
Om\left(x\right)=\frac{h\left(x\right)^2-1}{x^3-1},
   \end{equation}
where $h\left(x\right)=H\left(x\right)/H_0$ and $x=ln\left(z+1\right)^{-1}$. The mentioned discussion of the diagnostic tools has been made for understanding the behavior of a new dark energy model, but it cannot give us any advantageous information about the situation of stability of the model. From this, by employing the squared sound speed $v^2_s$ \cite{56}, checking the stability of the models against perturbations of the background will be achievable.\\

 In this paper, motivated from aforementioned cases we would like to study a new
model of HDE (NHDE) based on DGP braneworld with consideration of a non-gravitational
interaction between dark energy and dark matter. We investigate the behavior of present model in the context of the deceleration parameter and the equation of state. We also use the StateFinder pair and $Om$-diagnostic tool for investigation of the new HDE presented in this work. Moreover, we test the stability of the present models using the squared sound speed.
In particular, the analysis of the models with the help of $\chi^{2}_{min}$ and Markov chain Monte Carlo (MCMC) method using Pantheon, BAO and CMB observational data is performed. We can see that the present interactions are compatible with observations and make stable models in order to investigation of dark energy behavior. We can also see that the phantom behavior is accessible in these models. 
The structure of this paper is as follows. In the next section, we introduce the New Holographic Dark Energy model (NHDE). In section III, we present four phenomenological interactions to reach the proper terms for Hubble and dark energy for checking the evolution of the Universe. In section IV, employing the $Om$-diagnostic tool and the StateFinder pair, we investigate characteristics of the models. In section V, we extend the study to check the stability of the models. Finally, in the section VI, using the latest observational data free parameters in four different models will be constrained and also the appropriate cosmological model will be selected using Akaike Information
Criterion (AIC) and Bayesian Information Criterion (BIC).
The last section is devoted to concluding and remarks.
    \section{Background Evolution}
It is well-known that a homogeneous and isotropic Friedmann-Robertson-Walker universe can be described by
     \begin{equation}
  ds^2=-dt^2+a^2\left(t\right)\left(\frac{dr^2}{1-kr^2}+r^2d\Omega^2\right),
   \end{equation}
in which $k=0,1,-1$ denote a flat, closed and open universe respectively. According to the equation above and calculation of \cite{62} the Friedmann equations in DGP braneworld can be written as
 \begin{equation}\label{frwold}
H^2+\frac{k}{a^2}=\left(\sqrt{\frac{\rho}{3M_P^2}+\frac{1}{4r^2_c}}+\frac{\epsilon}{2r_c}\right)^2,
\end{equation}
    where $M_P=\sqrt{8\pi G}$  is the reduced Planck mass, $G$ is the Newton constant and $\rho=\rho_D+\rho_m$. In the consideration of limitation $r_c\rightarrow\infty$, the ordinary Friedmann equation is recovered.  In the modern cosmology, by the use of modern observations, we know that the Universe is spatially flat. Hence, a flat Friedmann-Robertson-Walker equation for $r_c\gg1$ can be written as
     \begin{equation}\label{frw1}
  H^2-\frac{\epsilon}{r_c}H=\frac{\rho}{3M_P^2},
   \end{equation}
   where $r_c=\frac{M^2_p}{2M^3_5}=\frac{G_5}{2G_4}$ stands for the crossover length scale determining the transition from 4D to 5D behavior, $\epsilon=\pm1$ corresponds to the two branches (self-accelerated and normal) of solution\cite{29}.  The  $\epsilon=+1$ is related to the self-accelerating solution in which the universe may enter an accelerating phase in the late time without additional dark energy component. The $\epsilon=-1$ corresponds to the universe which is accelerated provided that the dark energy component is set on the brane. Using these concepts of the DGP braneworld model, many authors have analyzed the physical behavior of the universe in order to constrain the cosmic parameters and check the changes in different models of dark energy \cite{108},  \cite{109}, \cite{79},\cite{80},\cite{81},\cite{82},\cite{83},\cite{84}.
  Assuming the following terms
   \begin{equation}\label{dens}
  \Omega_m=\frac{\rho_m}{3M_p^2H^2},~~~~~~~\Omega_D=\frac{\rho_D}{3M_p^2H^2},~~~~~~~\Omega_{r_c}=\frac{1}{4r_c^2H^2},
   \end{equation}
   the Eq. \ref{frw1} changes as follows
    \begin{equation}\label{frw2}
  \Omega_m+\Omega_D=1-2\epsilon\sqrt{\Omega_{r_c}} .
   \end{equation}
The effect of limit consideration of $r_c$ in comparison with Hubble scale can be found in the dimensionless parameter $\Omega_{rc}$ Eq. \ref{dens} similar to $\Omega_D$ and $\Omega_m$ and even $\Omega_k=-K/H_0^2$. As it mentioned for $r_c \rightarrow \infty$ the Eq. \ref{frw1} tend to standard cosmology. This component tied with the amount of dark energy and dark matter in the universe in this framework as the following term
\begin{equation}\label{crossover}
\Omega_D+\Omega_m+2\sqrt{\Omega_{rc}}=1,
\end{equation}
in which $\Omega_{rc}=\left(4r_c^2H^2\right)^{-1}$. 
The modified form of Eqs. \ref{frwold} and \ref{frw1} would be 
 \begin{equation}\label{modifiedfrw}
  \Omega_k+\left(\sqrt{\Omega_{rc}+\Omega}+\sqrt{\Omega_{rc}}\right)^2=1,
   \end{equation}
 in which $\Omega=\frac{8\pi G\rho}{3H^2}$ and one can study the behavior of the model under the existence of the curvature.  The energy density of the new holographic dark energy (NHDE) is given by the following relation\cite{29}
\begin{equation}\label{rhod}
   \rho_D=\frac{3c^2}{ L^2}\left(1-\frac{\epsilon L}{3 r_c}\right),~~~~~~~~~~  \Omega_D=c^2\left(1-\frac{2\epsilon\sqrt{\Omega_{r_c}}}{3}\right),
   \end{equation}
 where $L=H^{-1}$ is the Hubble horizon as the system's IR cutoff. In what follows, by this choice for the system's IR cutoff and constraining the present model by use of the latest observational data, we study the evolution of equation of state (EoS) and the deceleration parameter. We also survey the physical aspects of the current model by the use of two diagnostic tools known as Om-Diagnostic and StateFinder pair and we examine the stability of the model.
   \section{ Interacting NHDE}
  \justify
In this section, following recent work \cite{61} we would like to introduce the forms of non-gravitational interactions considered in this paper. But before, we would like to mention, that in modern cosmology, the non-gravitational interaction between any kind of dark energy and dark matter is understood in the following way
\begin{equation}\label{drhom}
\dot{\rho}_m+3H\rho_m=Q,
\end{equation}
\begin{equation}\label{drhod}
\dot{\rho}_D+3H\left(\rho_D+P_D\right)=-Q,
\end{equation}
which $Q$ is the interaction term, $b$ is the coupling constant, H is the Hubble parameter, $\rho_D$ is the density of dark energy of the present work (NHDE) and $\rho_m$ is the density of dark matter. In spite of the most common choice for $Q$-term as $3H(b_1\rho_D+b_2\rho_m)$ which $b_{1,2}$ are the coupling constant, it is more appropriate to use a single coupling constant (e.g. $Q_1=3Hb\rho_D$, $Q_2=3Hb\rho_m$ and $Q_3=3Hb(\rho_D+\rho_m)$. In our recent work, we compared different phenomenological interaction models including linear and nonlinear cases in the framework of the holographic ricci dark energy model (for more details see Ref.\cite{61.1}) and we found that the linear interaction $Q=3Hb\rho_D$ is the best case among the others. Hence, in this work for comparison between non-interacting (NHDE) and interacting (INHDE) form of the new holographic dark energy model we take $Q=3Hb\rho_D$ in our calculations.\\
To simplify future calculations and due to the consideration only different types of interaction it is reasonable to obtain some mathematical pattern, which can be used for any form of interaction term. In this regards, taking time derivative of Eq. \ref{frw1} and using Eqs. \ref{frw1}, \ref{drhom} and \ref{drhod} yields
  \begin{equation}\label{pd}
  P_D=-\frac{2}{3}\frac{\dot{H}}{H^2}\rho_m-\frac{\epsilon M_P^2}{r_c}\frac{\dot{H}}{H}-\rho_m.
   \end{equation}
Combining the Eqs. \ref{dens}, \ref{frw2} \ref{rhod}, \ref{drhod} and \ref{pd} we have
 \begin{equation}\label{final}
\dot{\Omega}_D+\left(2\Omega_D-2-3\left(\frac{\Omega_D-c^2}{c^2}\right)\right)\frac{\dot{H}}{H}+3\left(-1+2\Omega_D-3\left(\frac{\Omega_D-c^2}{c^2}\right)+\frac{\Omega_i}{3}\right)H=0,
   \end{equation}
  in which $\Omega_i=Q(3M_P^2H^3)^{-1}$. Now, for scrutinizing the evolution of the universe using Eqs. \ref{rhod} and \ref{final} we have the following two differential equations
     \begin{equation}\label{hz}
  \frac{dH\left(z\right)}{dz}=\frac{H\left(z\right)}{\left(1+z\right)}\left(\frac{6+3\Omega_D-9\left(\frac{\Omega_D}{c^2}\right)-\Omega_i}{1+c^2+\Omega_D\left(1-\frac{3}{c^2}\right)}\right),
   \end{equation}
     \begin{equation}\label{omegaz}
  \frac{d\Omega_D\left(z\right)}{dz}=\frac{\left(c^2-\Omega_D\right)}{\left(1+z\right)}\left(\frac{\left(6+3\Omega_D-9\left(\frac{\Omega_D}{c^2}\right)-\Omega_i\right)}{\left(1+c^2+\Omega_D\left(1-\frac{3}{c^2}\right)\right)}\right).
   \end{equation}
 For clarification of the calculations in the next parts one can write the Eq. \ref{hz}
   \begin{equation}\label{hz2}
  \frac{\dot{H}}{H^2}=\left(\frac{-6-3\Omega_D+9\left(\frac{\Omega_D}{c^2}\right)+\Omega_i}{1+c^2+\Omega_D\left(1-\frac{3}{c^2}\right)}\right).
   \end{equation}
These equations explain the behavior of Hubble parameter and the energy density of NHDE and will be used in the calculations by solving numerically.
\section{State of the Universe}
In this section, we would like to present and discuss the behavior of the models using the deceleration parameter and the equation of state. In order to simplify the discussion we have organized two subsections namely, the deceleration parameter and the equation of state.
  \subsection{ The deceleration parameter}
The deceleration parameter is defined by
\begin{equation}\label{dp}
q=-\frac{\ddot{a}a}{\dot{a}^2}=-1-\frac{\dot{H}}{H^2},
\end{equation}
where $a$ is the scale factor of the universe, H is the Hubble parameter and dots indicate the time derivative. The expansion of the universe will be accelerated if $\ddot{a}>0$ and in this case the deceleration parameter turns to be negative. Using the Eq. \ref{hz2} we find
\begin{equation}\label{dpt}
 q=-1-\left(\frac{-6-3\Omega_D+9\left({\Omega_D}{c^2}\right)+\Omega_i}{1+c^2+\Omega_D\left(1-\frac{3}{c^2}\right)}\right).
   \end{equation}
\begin{figure}[ht!]
   \centering
\begin{tabular}{cc}
\hspace*{-0.1in}
\includegraphics[width=0.45\textwidth]{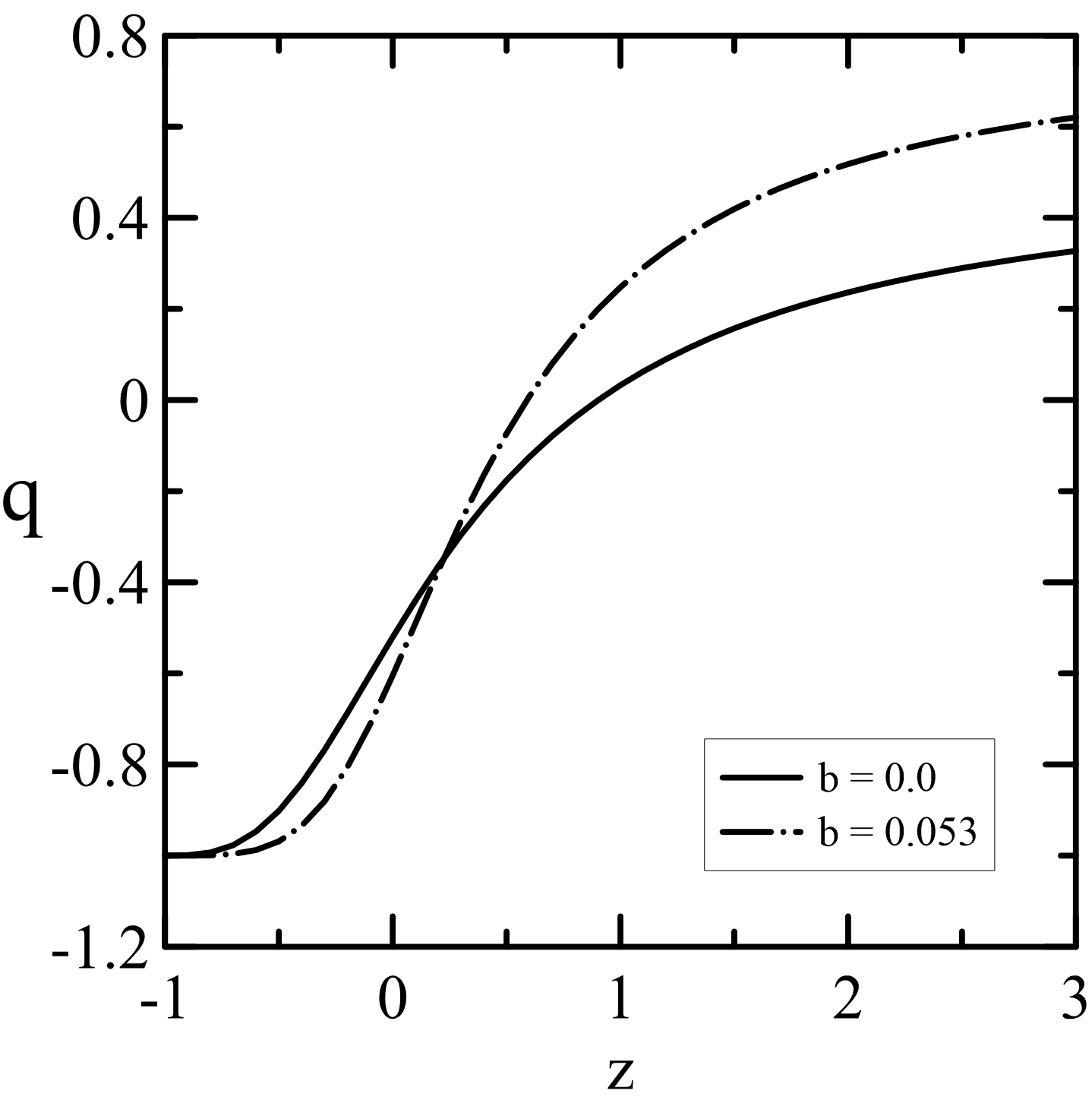}\hspace{5mm}
\includegraphics[width=0.45\textwidth]{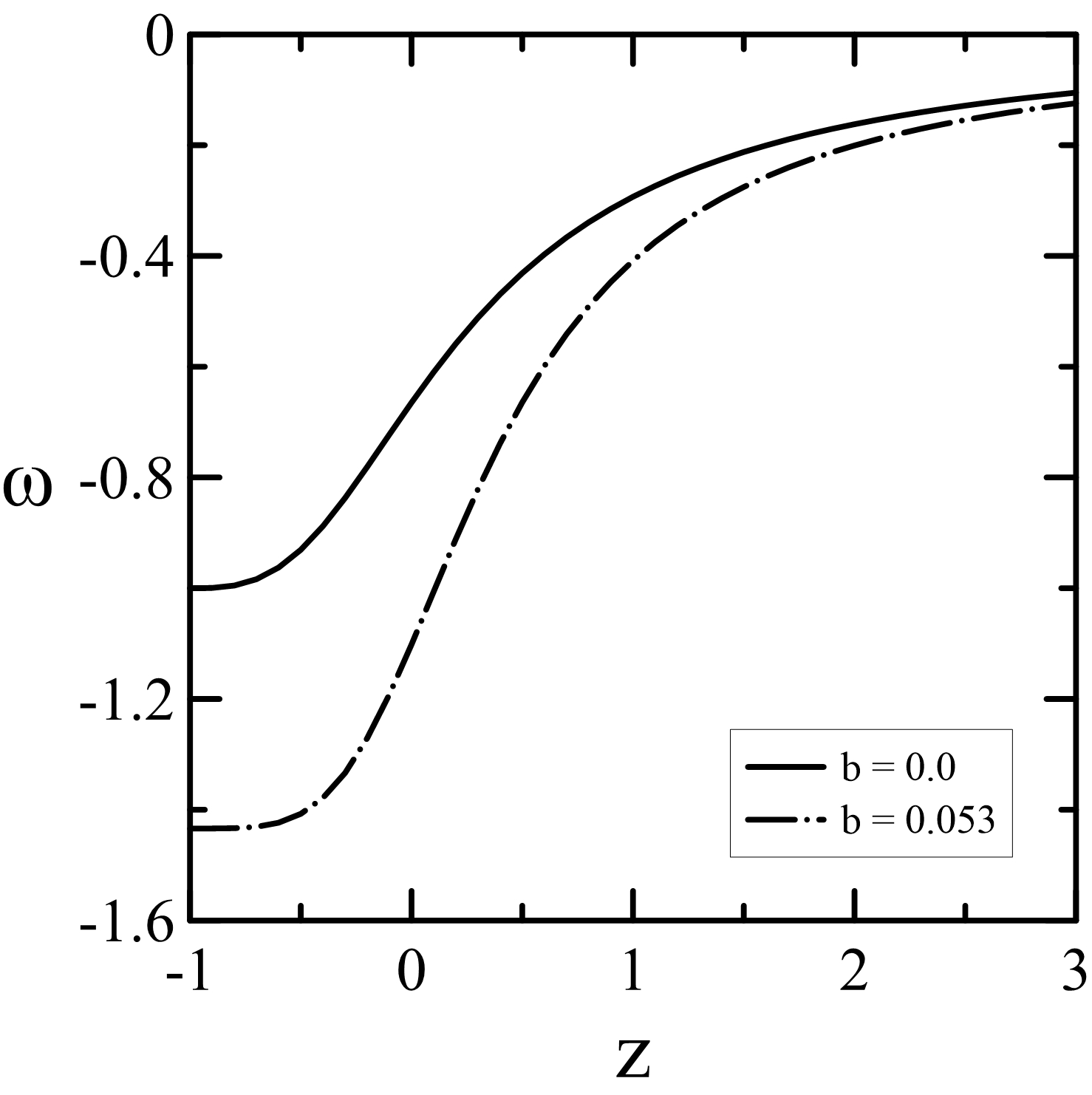}
\end{tabular}
\caption{\small The evolution of the deceleration parameter (left plane) and the equation of state (right plane) in terms of redshift. Dashed line indicates the interacting ($Q=3Hb\rho_D$) and solid line indicates the non-interacting model according to the best fitted value of parameters inserted in the Table\ref{bfs}.}\label{q}
\end{figure}
In Fig. \ref{q}we plotted the deceleration parameter the interacting and non-interacting form of the model. Using the deceleration parameter we can find the time of shifting from decelerating to accelerating universe\cite{100}. The interacting and non-interacting model due to the imposing an interaction between dark energy and dark matter show an accelerating universe $q<0$ shifting from matter dominated to dark energy dominated era. Observations suggest that the transition point from decelerating to accelerating time in the redshift range of $z\approx0.6$ and in range of $0.45<z<1$ \cite{{72.1},{72.2},{72.3},{72.4},{72.5},{99},{59},{60},{84},{29},{97},{98}} and also interacting model with ($q_0=-0.58$) and non-interacting model with($q_0=-0.51$) has a good agreement with the Planck value of the deceleration parameter ($q_0=-0.55$)\cite{39}. 
\subsection{The equation of state}
Taking time drivative of Eq. \ref{rhod} and using Eqs. \ref{drhod} and \ref{hz} we can study the evolution of EoS
   \begin{equation}\label{eos}
  \omega_D=-1-\frac{1}{3\Omega_D}\left(\Omega_i+\left(c^2-\Omega_D\right)\frac{\dot{H}}{H^2}-2\Omega_D\frac{\dot{H}}{H^2}\right),
   \end{equation}
 Using the Eq. \ref{hz2} we easily reach the following term
\begin{equation}\label{eost}
 \omega_D=-1-\frac{1}{3\Omega_D}\left(\Omega_i+\left(c^2-\Omega_D\right)\left(\frac{-6-3\Omega_D+9\left(\frac{\Omega_D}{c^2}\right)+\Omega_i}{1+c^2+\Omega_D\left(1-\frac{3}{c^2}\right)}\right)-2\Omega_D\left(\frac{-6-3\Omega_D+9\left(\frac{\Omega_D}{c^2}\right)+\Omega_i}{1+c^2+\Omega_D\left(1-\frac{3}{c^2}\right)}\right)\right).
\end{equation}
Now we provide the easy use non-interacting and interacting equation of state of the model
\begin{equation}\label{eosnob}
  \omega_D=\frac{2\left(c^2-\Omega_D\right)}{\Omega_D\left(1+c^2+\Omega_D\left(1-\frac{3}{c^2}\right)\right)}.
   \end{equation}
 \begin{equation}\label{eosb}
  \omega_D=\frac{2\left(c^2-\Omega_D\right)+b\Omega_D\left(3\Omega_D-c^2\right)}{\Omega_D\left(1+c^2+\Omega_D\left(1-\frac{3}{c^2}\right)\right)},
   \end{equation}
respectively. Regarding to the Fig. \ref{q} we can discuss the equation of state for the interacting and non-interacting model. As this figure shows, according to the relation between the deceleration parameter and the equation of state, we can see that the trajectories of the equation of state for both models cross the line of $\omega_D=-0.33$ in redshift range $0.45<z<1$ \cite{{72.1},{72.2},{72.3},{72.4},{72.5},{99},{59},{60},{84},{29},{97},{98}}. Of course, the interacting model has the ability of crossing the phantom divided line $\omega_D=-1$ at the late time $z<0$. In this figure, it can be observed that the present model (NHDE) has the behavior similar to $\Lambda$CDM model for non-interacting model.
\section{ Diagnostic recognition}
In this section we are going to present and discuss the behavior of the models using StateFinder and $Om$ analysis. In order to simplify future discussion we have organized two subsections namely, the StateFinder pair and the $Om$-diagnostic tool.
\subsection{ The StateFinder pair}
\begin{figure}[H]
\centering
\begin{tabular}{ccc}
\hspace*{-0.1in}
\includegraphics[width=0.50\textwidth]{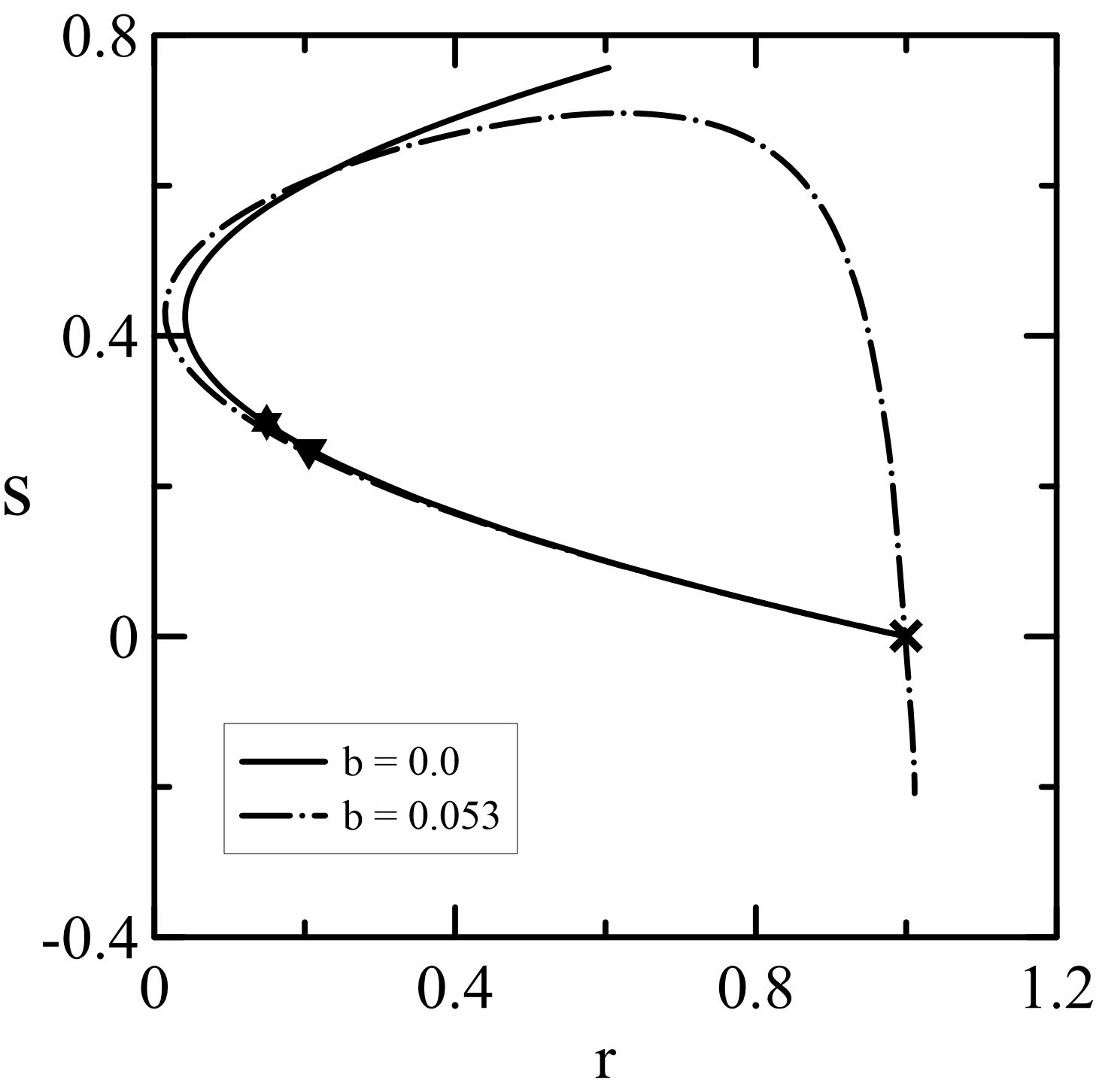}\end{tabular}
\caption{\small The evolution of parameter $s$ in terms of parameter $r$. Dashed line indicates the interacting ($Q=3Hb\rho_D$) and solid line indicates the non-interacting model according to the best fitted value of parameters inserted in the Table \ref{bfs}. The cross symbol denotes the $\Lambda$CDM model star symbol represents the present value of non-interacting model and the triangle symbol represents the present value of interacting model.} \label{sr}
\end{figure}

Using the q. \ref{r-s} we have plotted the StateFinder pair (s in terms of r) in Fig. \ref{sr}.
For $\left(\frac{\ddot{H}}{H^3}-2\frac{\dot{H}}{H^2}\right)$ in Eq. \ref{r-s} by taking the time derivative of both sides of Eq. \ref{hz} we have
\begin{equation}\label{ddhz}
\begin{split}
\frac{\ddot{H}}{H^3}=\left(-\left(3\Omega'_D\left(\frac{3}{c^2}-1\right)+3b\Omega'_i\right)\left(\Omega_D\left(\frac{3}{c^2}-1\right)-c^2-1\right)+3\Omega'_D\left(\frac{3}{c^2}-1\right)\left(1+\Omega_D\left(\frac{3}{c^2}-1\right)-3+3b\Omega'_D\right)\right)\\
\times\left(\Omega_D\left(\frac{3}{c^2}-1\right)-c^2-1\right)^{-2}+2\left(\frac{\dot{H}}{H^2}\right)^2.~~~~~~~~~~~~~~~~~~~~~~~~~~~~~~~~~~~~~~~~~~~~~~~~~~~~~~~~~~~~~~~~~~~~~~~~~~~~~~
\end{split}
\end{equation}
in which $\Omega'_D=\frac{\dot{\Omega}_D}{H}$ and taking time derivative of Eq. \ref{rhod} yields
\begin{equation}\label{Omegad}
\dot{\Omega}_D=\frac{2\epsilon c^2 \sqrt{\Omega_{rc}}}{3}\frac{\dot{H}}{H}.
\end{equation}
It can be seen that for non-interacting models as the universe expands, by increasing the value of parameter $r$, the parameter $s$ moves from positive to negative values while the non-interacting model start the movement from zero point to the positive area and again back to the negative region. The fixed point $ \left(r, s\right) = \left(1,0\right)$ represents the $\Lambda$CDM scenario. Checking the track of each case shows us that interacting model has both the Chaplygin gas behavior $\left(s<0, r>1\right)$ and the quintessence behavior $\left(s>0,r<1\right)$ and non-interacting model behave similar quintessence. Of course the trajectories of both models meet the fixed point $\left(1,0\right)$ indicating the evolution from quintessence to phantom-like behavior as the universe expands. The behavior of interacting model also in comparison to the non-interacting form is close to $\Lambda$CDM. Moreover, for simple power law evolution of the scale factor $a\left(t\right)\approx t^{0.66\alpha}$, it can be easily found $r=\left(1-3\alpha\right)\left(1-1.5\alpha\right)$ and $s=\alpha$ \cite{107}. Accordingly, $s<0$ corresponds to a phantom-like dark energy appearing in the non-interacting model. This is an affirmation on the equation of state results.
\subsection{ The $Om$-diagnostic tool}
Fig. \ref{om} shows the $Om$-diagnostic trajectories the interacting and non-interacting models. The advantage of the $Om$-diagnostic is its less dependency on the matter density relative to the equation of state of dark energy.
In this figure we can analyze the results according to the area under the trajectories. If the value of trajectories are in the positive, null and negative region it can correspond to the phantom ($\omega<-1$), $\Lambda$CDM ($\omega=-1$) and quintessence ($\omega>-1$), respectively. Hence we cans see that at the late time models show the phantom-like behavior and for values of bigger redshift, both models are in the quintessence area.
\begin{figure}[H]
   \centering
\begin{tabular}{ccc}
\hspace*{-0.1in}
\includegraphics[width=0.5\textwidth]{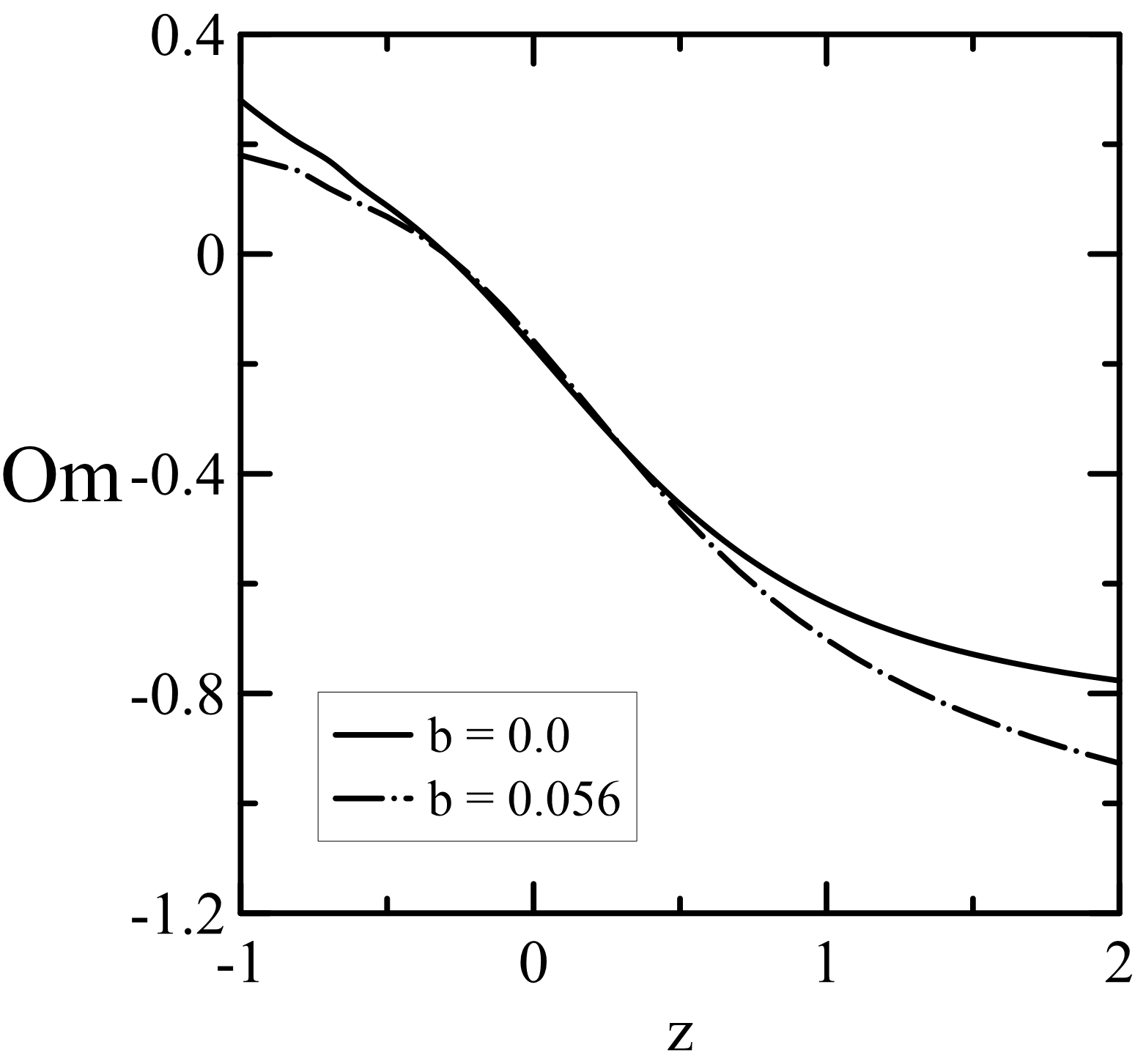}\end{tabular}
\caption{\small The evolution of the $Om$-diagnostic tool in terms of redshift. Dashed line indicates the interacting ($Q=3Hb\rho_D$) and solid line indicates the non-interacting model according to the best fitted value of parameters inserted in the Table \ref{bfs}.} \label{om}
\end{figure}
\section{Stability}
\begin{figure}[H]
   \centering
\begin{tabular}{ccc}
\hspace*{-0.1in}
\includegraphics[width=0.5\textwidth]{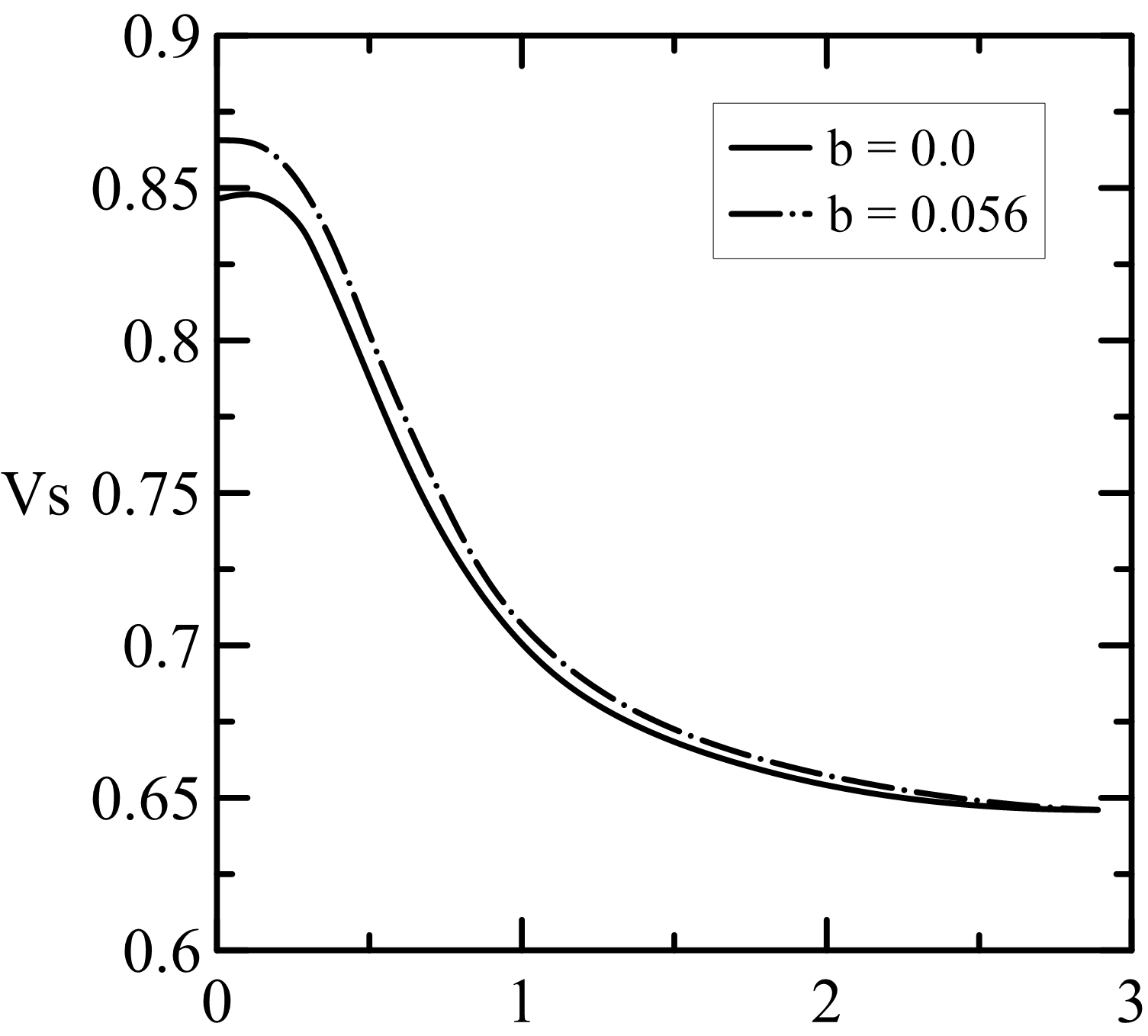}\end{tabular}
\caption{\small The evolution of $v^2_s$ versus redshift. Dashed line indicates the interacting ($Q=3Hb\rho_D$) and solid line indicates the non-interacting model according to the best fitted value of parameters inserted in the Table \ref{bfs}. The positive value of trajectory for each model shows the stability against perturbation of the background.} \label{vsimg}
\end{figure}

In order to test viability of a new dark energy model we refer to investigate the stability of the model against perturbation. The behavior of square sound speed ($v^2_s$) \cite{56} can be studied as an approach to check the stability of a new dark energy model. It is claimed that the sign of $v_s^2$ is important to specify the stability of background evolution. The signs of squared sound speed $v^2_s>0$ and $v^2_s<0$ denote a stable and instable universe against perturbation respectively. The perturbed energy density of the background in a linear perturbation structure is
\begin{equation}\label{ped}
\rho\left(x,t\right)=\rho\left(t\right)+\delta\rho\left(x,t\right),
\end{equation}
in which $\rho\left(t\right)$ is unperturbed energy density of the background. The equation of energy conservation is \cite{56}
\begin{equation}\label{ece}
\delta\ddot{\rho}=v^2_s\bigtriangledown^2\delta\rho\left(x,t\right).
\end{equation}
For positive sign of squared sound speed the Eq. \ref{ece} will be a regular wave equation which its solution can be obtained as $\delta\rho=\delta\rho_0e^{-i\omega_0t+ikx}$ indicating a propagation state for density perturbation. It is easy to see that the squared sound speed can be written as
\begin{equation}\label{vs}
v^2_s=\frac{\dot{P}}{\dot{\rho}}=\frac{\dot{\omega}_D\rho}{\dot{\rho}}+\omega_D,
\end{equation}
taking time derivative of Eqs. \ref{rhod} and \ref{eos} and combining with Eqs.\ref{ddhz} and \ref{vs} one can plot the evolution of $v^2_s$ in terms of redshift as it is shown in Fig.\ref{vsimg}. During the cosmic evolution, the both interacting and non-interacting models in comparison with GDE\cite{45} \cite{58}, SMHDE\cite{44}, ADE \cite{57} and also HDE in the standard cosmology which are instable against perturbations\cite{43} show stability against background perturbations in early time, present and late time.

\begin{table}[H]
\centering
\caption{The fitted values of cosmological parameters for the interacting and non-interacting new holographic dark energy model. We also provide the fitted parameter of $\Lambda$CDM and holographic ricci dark energy model for having more accurate comparison with other models. The AIC and BIC stand for Akaike Information Criterion and Bayesian Information Criterion respectively. Also, $\Delta AIC=AIC_i-AIC_{min}$ in which $i$ denotes the number of models $\{i=1,2,...,N\}$.}
\label{bfs}
\hspace*{-1.5em}
\footnotesize\addtolength{\tabcolsep}{3pt}
\begin{tabular}{ccccccc}
\hline\hline
\multicolumn{5}{c}{ }                                             \\ [-0.4cm]
\multicolumn{5}{c}{ Linear Interactions}                                             \\ [-0.02cm] \hline
\multicolumn{1}{c}{Params}                 & \multicolumn{1}{c}{$\Lambda$CDM}    & \multicolumn{1}{c}{$HRDE$}          & \multicolumn{1}{c}{$NHDE(N/A)$}        & \multicolumn{1}{c}{$NHDE(\rho_D)$}  \\ [-0.02cm]\hline
\multicolumn{1}{c}{}                              & \multicolumn{1}{l}{}    & \multicolumn{1}{l}{}                    & \multicolumn{1}{l}{}                       & \multicolumn{1}{l}{}	        \\  [-0.2cm]
\multicolumn{1}{c}{$H_0$ }   & \multicolumn{1}{c}{$68.8878_{-3.5762}^{+4.0012}$} & \multicolumn{1}{c}{$68.972 _ { - 3.7599 } ^ { + 3.6415 }$}   & \multicolumn{1}{c}{$68.8091_{-4.2458}^{+2.6597}$}   & \multicolumn{1}{c}{$68.8660_{- 4.3027}^{+2.6028}$}  \\ [0.15cm]
\multicolumn{1}{c}{$\Omega_D$ }             & \multicolumn{1}{c}{$0.7220_{-0.08982}^{+0.8050}$} & \multicolumn{1}{c}{$0.6965 _ { - 0.0851 } ^ { + 0.1150 }$}   & \multicolumn{1}{c}{$0.6923_{-0.0164}^{+0.0137}$}    & \multicolumn{1}{c}{$0.6890_{-0.0124}^{+0.0121}$}  \\ [0.15cm]
\multicolumn{1}{c}{$c$}      & \multicolumn{1}{c}{$-$}  & \multicolumn{1}{c}{$ 0.4240 _ { - 0.1141 } ^ { + 0.1864 }$}     & \multicolumn{1}{c}{$0.8709_{-0.0095}^{+0.0072}$}     & \multicolumn{1}{c}{$0.8780_{-0.0018}^{+0.0021}$}  \\ [0.15cm]
\multicolumn{1}{c}{$b$}      & \multicolumn{1}{c}{$-$} & \multicolumn{1}{c}{$-$}     & \multicolumn{1}{c}{$-$}     & \multicolumn{1}{c}{$0.0535_{-0.0339}^{+0.0398}$}\\ [0.15cm]
\multicolumn{1}{c}{$M$}      & \multicolumn{1}{c}{$-19.3867 _ { - 0.0209 } ^ { + 0.0206 }$} & \multicolumn{1}{c}{$-19.3846 _ { - 0.0948 } ^ { + 0.0778 }$}     & \multicolumn{1}{c}{$-19.3799_{-0.1106}^{+0.078}$}     & \multicolumn{1}{c}{$-19.3865_{-0.104}^{+0.0846}$}  \\ [0.15cm]
\multicolumn{1}{c}{$\chi^2$}      & \multicolumn{1}{c}{$1030.6021$} & \multicolumn{1}{c}{$1031.9125$}     & \multicolumn{1}{c}{$1034.8899$}     & \multicolumn{1}{c}{$1032.1231$} \\  [0.15cm]
\multicolumn{1}{c}{$\chi_{dof}$}      & \multicolumn{1}{c}{$0.9732$} & \multicolumn{1}{c}{$0.9744$}     & \multicolumn{1}{c}{$0.9772$}     & \multicolumn{1}{c}{$0.9746$} \\  [0.15cm] 
\multicolumn{1}{c}{$AIC$}      & \multicolumn{1}{c}{$1036.6021$} & \multicolumn{1}{c}{$1039.9126$}     & \multicolumn{1}{c}{$1042.8899$}     & \multicolumn{1}{c}{$1042.1232$} \\  [0.15cm] 
\multicolumn{1}{c}{$\Delta AIC$}      & \multicolumn{1}{c}{$0$} & \multicolumn{1}{c}{$3.3105$}     & \multicolumn{1}{c}{$6.2878$}     & \multicolumn{1}{c}{$5.5210$} \\  [0.15cm] 
\multicolumn{1}{c}{$BIC$}      & \multicolumn{1}{c}{$1051.4974$} & \multicolumn{1}{c}{$1059.7729$}     & \multicolumn{1}{c}{$1062.7502$}     & \multicolumn{1}{c}{$1066.9486$} \\  [0.15cm] 
\multicolumn{1}{c}{$\Delta BIC$}      & \multicolumn{1}{c}{$0$} & \multicolumn{1}{c}{$8.2755$}     & \multicolumn{1}{c}{$11.2453$}     & \multicolumn{1}{c}{$15.4417$} \\  [0.15cm] \hline\hline

\end{tabular}
\end{table}

\section{Data Analysis Methods}
To analyze the models and to obtain the best fit values for the model parameters; in this paper we combine the latest observational data including SNIa, BAO and CMB. For this purpose, we employed the public codes EMCEE \cite{39.1} and GetDist Python package\footnote{https://getdist.readthedocs.io} for analyzing and plotting the contours. in order to fit the cosmological parameters for $1\sigma$ and $2\sigma$ confidence area. This method also provides reliable error estimates on the measured variables.\\
For supernova, we use 1048 data points of the recent proposed Pantheon data \citep{40.1}. We use the systematic covariance $C_{sys}$ for a vector of binned distances
\begin{equation}\label{SNsys}
C_{ij,sys}=\sum_{n=1}^{i}\left(\frac{\partial \mu_i}{\partial S_n}\right)\left(\frac{\partial \mu_j}{\partial S_n}\right)\left(\sigma_{S_k}\right)
\end{equation}
in which the summation is over the $n$ systematics with $S_n$ and its magnitude of its error $\sigma_{S_n}$. According to $\triangle\mu=\mu_{data}-M-\mu_{obs}$ in which $M$ is a nuisance parameter we can write the $\chi^2$ relation for Pantheon SNIa data as
\begin{equation}\label{SNchi2}
\chi^2_{Pantheon}=\triangle\mu^T\cdot C_{Pantheon}^{-1}\cdot\triangle\mu
\end{equation}
Note that the $C_{Pantheon}$ is the summation of the systematic covariance and statistical matrix $D_{stat}$ having a diagonal component. The complete version of full and binned Pantheon supernova data can be found in the online source\footnote{https://archive.stsci.edu/prepds/ps1cosmo/index.html}\\
We combine the extended Baryon Oscillation Spectroscopic Survey (eBOSS) quasar clustering at $z=1.52$ \citep{40.2}, isotropic BAO measurements of 6dF survey at an effective redshift ($z=0.106$) \citep{40.3} and the BOSS DR12 \cite{40} including six data points of Baryon Oscillations as the latest observational data for BAO. The $\chi^2_{BAO}$ of BOSS DR12 can be explained as
\begin{equation}\label{BOSSDR12}
\chi^2_{BOSS~DR12}=X^tC_{BAO}^{-1}X,
\end{equation}
where $X$ for six data points is
\begin{equation}\label{XBAO}
X=\left(\begin{array}{c} \frac{D_M\left(0.38\right)r_{s,fid}}{r_s\left(z_d\right)}-1512.39\\
\frac{H\left(0.38\right)r_s\left(z_d\right)}{r_s\left(z_d\right)}-81.208\\
\frac{D_M\left(0.51\right)r_{s,fid}}{r_s\left(z_d\right)}-1975.22\\
\frac{H\left(0.51\right)r_s\left(z_d\right)}{r_s\left(z_d\right)}-90.9\\
\frac{D_M\left(0.61\right)r_{s,fid}}{r_s\left(z_d\right)}-2306.68\\
\frac{H\left(0.51\right)r_s\left(z_d\right)}{r_s\left(z_d\right)}-98.964\end{array}\right),
\end{equation}
and $r_{s,fid}=$147.78 Mpc is the sound horizon of fiducial model, $D_M\left(z\right)=\left(1+z\right)D_A\left(z\right)$ is the comoving angular diameter distance. The sound horizon at the decoupling time $r_s\left(z_d\right)$ is defined as

\begin{equation}\label{BAO}
r_s\left(z_d\right)=\int_{z_d}^{\infty} \frac{c_s\left(z\right)}{H\left(z\right)}dz,
\end{equation}
in which $c_s=1/\sqrt{3\left(1+R_b/\left(1+z\right)\right)}$ is the sound speed with
$R_b=31500\Omega_bh^2\left(2.726/2.7\right)^{-4}$. The covariance matrix $Cov_{BAO}$ \cite{40} is:
\begin{equation}\label{iCOVBAO}
C^{-1}_{BAO}=\begin{pmatrix}   624.707& 23.729  &325.332    &8.34963&   157.386 &3.57778\\
23.729  &5.60873    &11.6429    &2.33996    &6.39263    &0.968056\\
325.332 &11.6429    &905.777    &29.3392    &515.271&   14.1013\\
8.34963 &2.33996    &29.3392    &5.42327    &16.1422&   2.85334\\
157.386 &6.39263    &515.271    &16.1422    &1375.12&   40.4327\\
3.57778 &0.968056   &14.1013    &2.85334    &40.4327
&6.25936\end{pmatrix}.
\end{equation}
The $\chi^2$ for combined data is
\begin{equation}\label{BAO}
\chi^2_{BAO}=\chi^2_{BOSS~DR12}+\chi^2_{6dF}+\chi^2_{eBOSS},
\end{equation}
Discovering the expansion history of the universe, we check Cosmic Microwave Background (CMB). For this, we use the data of Planck 2015 \cite{39}. The $\chi^2_{CMB}$ function may be explained as
\begin{equation}\label{CMB}
\chi^2_{CMB}=q_i-q^{data}_i Cov^{-1}_{CMB}\left(q_i,q_j\right),
\end{equation}
where $q_1=R\left(z_*\right)$, $q_2=l_A\left(z_*\right)$ and $q_3=\omega_b$ and $Cov_{CMB}$
is the covariance matrix \cite{39}. The data of Planck 2015 are
\begin{equation}\label{PLANCKDATA}
q^{data}_1=1.7382,~\\
q^{data}_2=301.63,~\\
q^{data}_3=0.02262.
\end{equation}
The acoustic scale $l_A$ is
\begin{equation}\label{lA}
l_A=\frac{3.14d_L\left(z_*\right)}{\left(1+z\right)r_s\left(z_*\right)},
\end{equation}
in which $r_s\left(z_*\right)$ is the comoving sound horizon at the drag epoch ($z_*$). The function of redshift at the drag epoch is \cite{104}

\begin{equation}\label{z_*}
z_*=1048\left[1+0.00124\left(\Omega_bh^2\right)^{-0.738}\right]\left[1+g_1\left(\Omega_mh^2\right)^{g_2}\right],
\end{equation}
where
\begin{equation}\label{g1 g2}
g_1=\frac{0.0783\left(\Omega_bh^2\right)^{-0.238}}{1+39.5\left(\Omega_bh^2\right)^{-0.763}}, ~~~g_2=\frac{0.560}{1+21.1\left(\Omega_bh^2\right)^{1.81}}.
\end{equation}
The CMB shift parameter is \cite{105}
\begin{equation}\label{R}
R=\sqrt{\Omega_{m_0}}\frac{H_0}{c}r_s\left(z_*\right).
\end{equation}
The reader should notice that the usage of CMB data does not provide the full Planck information but it is an optimum way of studying wide range of dark energy models.\\ 
The data for BAO and CMB could be found in the online source of latest version of MontePython \footnote[1]{http://baudren.github.io/montepython.html}.
Using minimized $\chi^2_{min}$, we can constrain and obtain the best-fit values of the free parameters. 
\begin{equation}\label{chi}
\chi_{min}^2=\chi_{SNIa}^2+\chi_{CMB}^2+\chi_{BAO}^2.
\end{equation}The best-fit values of $\Omega_D$, $ H_0$, $ \Omega_{rc}$, c, $b^2 $  and M by consideration of the $1\sigma$ and $2\sigma$ confidence level are shown in the Table \ref{bfs}.
Despite the fact that $\chi^2$ is known as the effective way of understanding the best values of free parameters, it cannot be only used to determine the best model between variety of models. Hence, for this issue Akaike Information Criterion (AIC) \cite{46} and Bayesian Information Criterion (BIC) \cite{47} have been proposed. For further information see \cite{48}, \cite{49}, \cite{50}, \cite{51}.
The AIC can be explained as
\begin{equation}\label{av}
AIC=-2\ln\mathcal{L}_{max}+ 2k,
\end{equation}
where $-2\ln\mathcal{L}_{max}=\chi^2_{min}$ is the highest likelihood, $k$ is the number of free parameters (2 for $\Lambda$CDM and 4 for NHDE models in addition of one further parameter $M$ for SNIa) and N is the number of data points used in the analysis. The BIC is similar to AIC with different second term
\begin{equation}\label{av}
BIC=-2\ln\mathcal{L}_{max}+ k\ln N.
\end{equation}
 It is obvious that a model favored by the observations should give a small AIC and a small BIC. Hence, we explain the levels of supporting the models from AIC and BIC.\\
 The level of support for each model from AIC is
\begin{itemize}
\item \textbf{Less than 2}: This indicates there is substantial evidence to support the model (i.e., the model can be considered almost as good as the best model).
\item\textbf{Between 4 and 7}: This indicates that the model has considerably less support.
\item  \textbf{Between 8 and 10 or bigger}: This indicates that there is essentially no support for the model (i.e., it is unlikely to be the best model).
\end{itemize}
The level depiction of evidence against models if the tool of selection is BIC:
\begin{itemize}
\item \textbf{Less than 2}:  It is not worth more than a bare mention (i.e., the model can be considered almost as good as the best model).
\item \textbf{Between 2 and 6}: The evidence against the model is positive.
\item \textbf{Between 6 and 10}: The evidence against the candidate model is strong.(i.e., it can be merely the best model).
\item \textbf{Bigger than 10}: The evidence is very strong (i.e., it is unlikely to be the best model).
\end{itemize}
\begin{figure}[H]
\begin{center}
\includegraphics[width=0.45\textwidth]{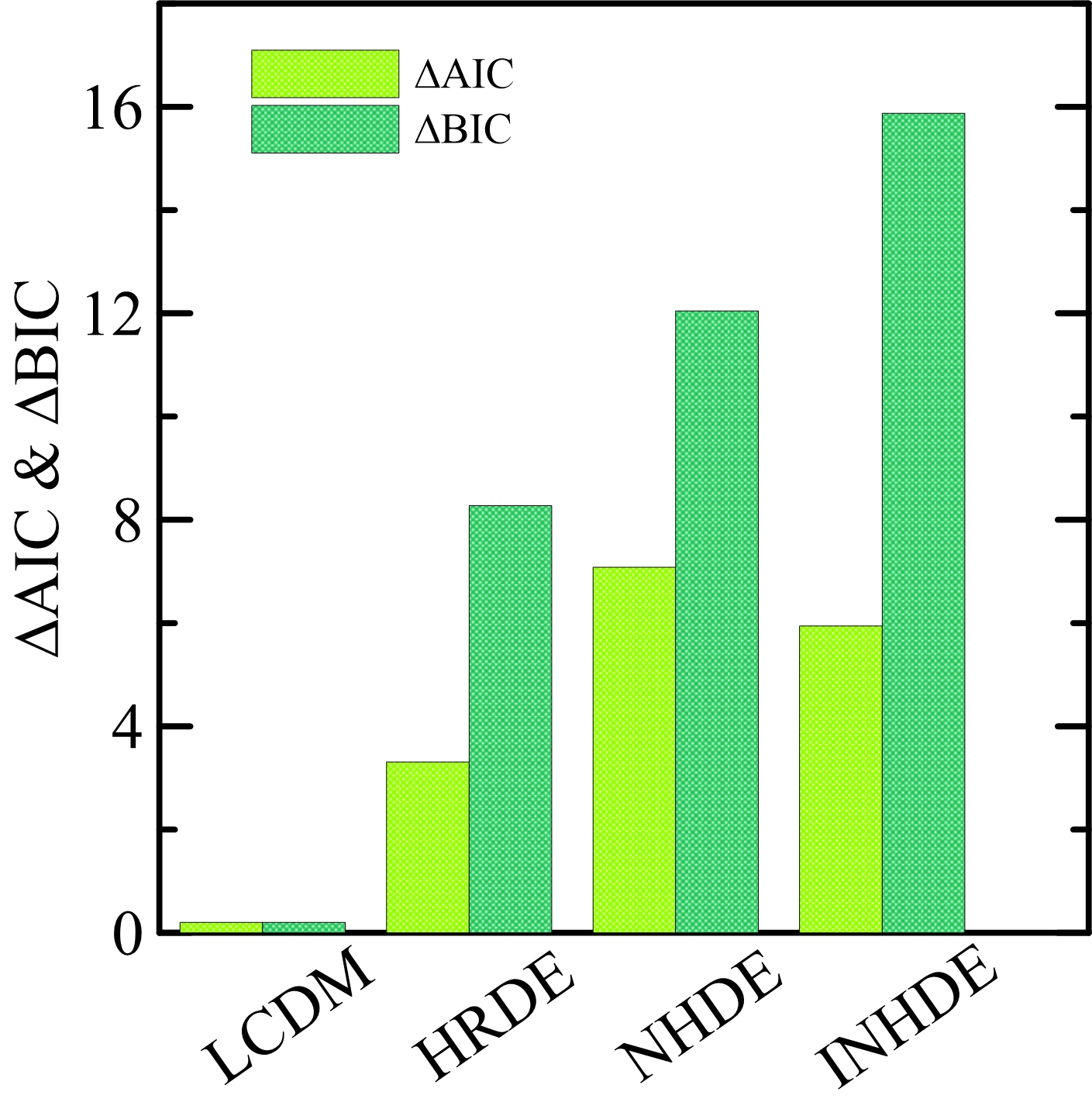}
\caption{Graphical presentment of $\Delta AIC$ and $\Delta BIC$} obtained from MCMC for all models [see Table \ref{bfs}].
\label{aicbic}
\end{center}
\end{figure}

Obviously, the value of $\chi^2$ for interacting model is smaller than the non-interacting one which is because of the additional parameter $b$. It can be seen that the value of interacting and non-interacting model are bigger than the $\Lambda$CDM and even holographic ricci dark energy model.\\
According to the AIC and BIC evidences shown in the Table\ref{bfs} and graphical representation of models comparison in Fig. \ref{aicbic} it is shown that by assumption of $\Lambda$CDM as the reference model the NHDE model cannot be supported by observational data and is ruled out by both AIC and BIC. Of course it should be noted that the proposing of the holographic dark energy models is a way to overcome the problems with which $\Lambda$CDM is faced. Thus, by consideration of RDE as the reference model the INHDE can be considered as a model which is favored by observational data only in the holographic area. The non-interacting model is considerably less supported by AIC and BIC shows the positive evidence against it compare to the interacting model.
   \begin{figure}[ht!]
\begin{center}
\includegraphics[width=0.8\textwidth]{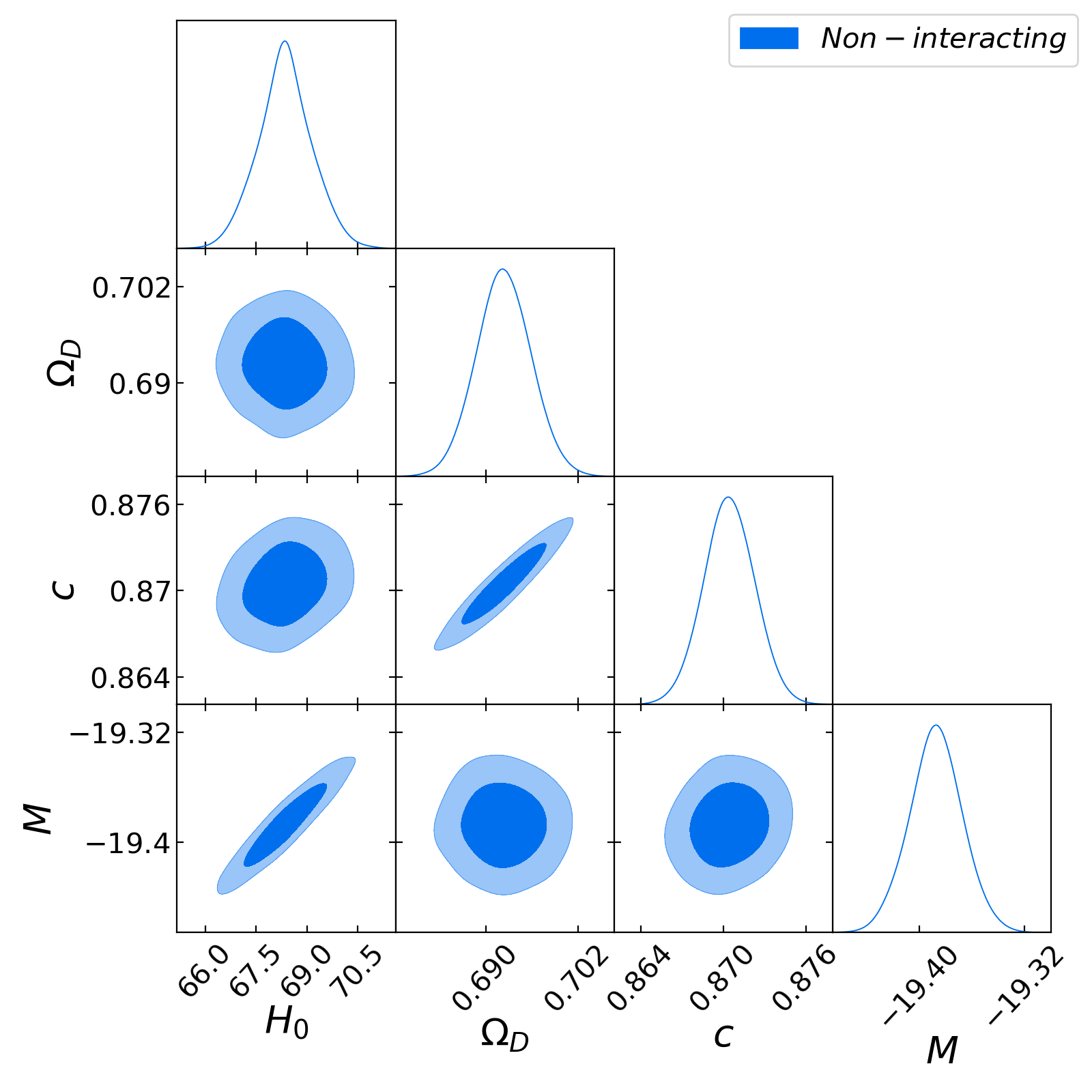}
\caption{The contour maps of $H_0$, $\Omega_D$ and $c$  for non-interacting model (NHDE) with $1\sigma\left(68.3\%\right)$ and  $2\sigma\left(95.4\%\right)$ confidence level.}
\label{t1}
\end{center}
\end{figure}
\begin{figure}[t]
\begin{center}
\includegraphics[width=0.85\textwidth]{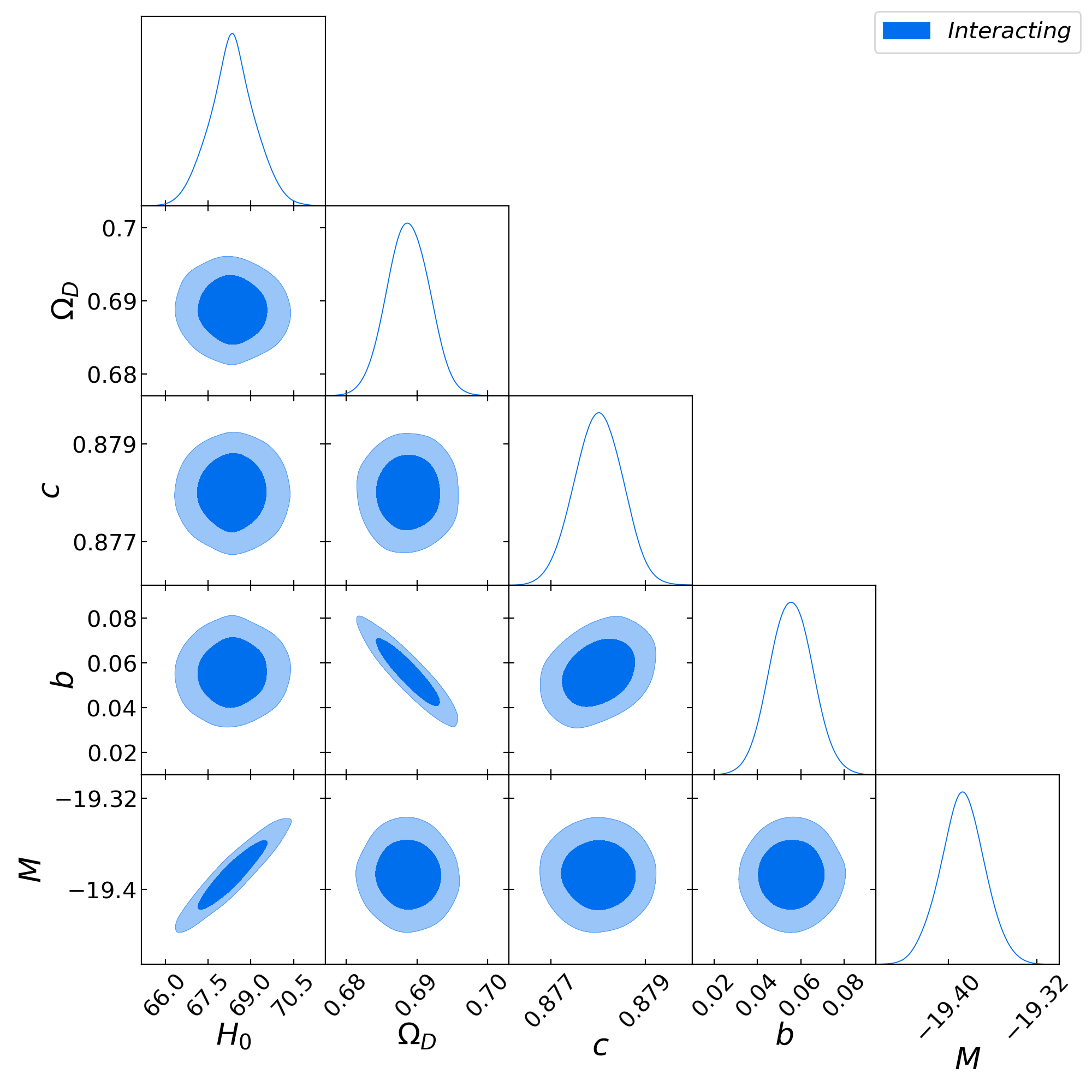}
\caption{The contour maps of $H_0$, $\Omega_D$, $c$ and $b$ for interacting model (INHDE) with $1\sigma\left(68.3\%\right)$ and  $2\sigma\left(95.4\%\right)$ confidence level.}
\label{t2}
\end{center}
\end{figure}
    \section*{ VIII.  CONCLUSION}
In the present work, we studied a New Holographic Dark Energy model (NHDE) with Hubble horizon as IR cutoff in the framework of the flat FRW with taking into account the non-gravitational interaction between dark matter and holographic dark energy ($Q=3bH\rho_D$). We used the latest observational data sets, namely Pantheon SNIa, Baryon Acoustic Oscillations (BAO) from BOSS DR12 and the Cosmic Microwave Background (CMB) of Planck 2015. We found that for both interacting and non-interacting models the corresponding universe is expanding and also accelerating. We found that the present value of the deceleration parameter has a good agreement with the Planck 2015 data and the transition redshift has a good compatibility with recent works on observational data ($0.4<z<1$). We found that the StateFinder trajectory for all models embrace $\Lambda$CDM model $(r,s)=(1,0)$ and also behave similar to the both quintessence and Chaplygin gas dark energy models. Using the $Om$-diagnostic tool by taking $ x=ln(z+1)^{-1}$, the evolution in terms of redshift shows positive values in the late time which implies the Phantom-like behavior and negative values for present and the early time denoting the quintessence behavior similar to the results of the equation of state. The interacting model, according to the StateFinder tool and the equation of state leads the phantom-like behavior which is one of the conditions of avoiding the creation of black hole's mass. For further investigation, we studied the stability of the considered models using the evolution of the squared sound speed $v^2_s$. In spite of the growth of background perturbations, the models show suitable stability. The mentioned results have been obtained using the fitted free parameters of the present model. We used MCMC method by employing EMCEE Python package. In order to study the compatibility of the models with observational data with the help of AIC and BIC criteria we found that the NHDE model in both interacting and non-interacting form is ruled out and is not favored by observational data. This result can be obtained once the $\Lambda$CDM is taken as the reference model. But according to this case that HDE models has been proposed for alleviation of $\Lambda$CDM problems, one can compare the NHDE with other HDE models the reference rather than the $\Lambda$CDM. Using this condition, the INHDE model can be considered as the compatible model with observational data but the non-interacting model still remains in less supporting area. In conclusion, the NHDE is compatible with behavior of the universe, but is cannot satisfies the condition of partiality from observational data.
      \section*{\centering \small ACKNOWLEDGMENTS}
 Martiros Khurshudyan is supported in part by a CAS President's International Fellowship Initiative Grant (No. 2018PM0054) and the NSFC (No. 11847226).
    \bibliographystyle{unsrt}

  \end{document}